# *Probing Superluminal Neutrinos Via Refraction*

**Albert Stebbins**
Fermilab Theoretical Astrophysics Group
stebbins@fnal.gov

## Abstract

One phenomenological explanation of superluminal propagation of neutrinos, which may have been observed by OPERA and MINOS, is that neutrinos travel faster inside of matter than in vacuum. If so neutrinos exhibit refraction inside matter and should exhibit other manifestations of refraction, such as deflection and reflection. Such refraction would be easily detectable through the momentum imparted to appropriately shaped refractive material inserted into the neutrino beam. For NuMI this could be as large as $\sim 10\,\mathrm{g\,cm/s}$. If these effect were found, they would provide new ways of manipulating and detecting neutrinos. Reasons why this scenario seems implausible are given, however it is still worthwhile to conduct simple searches for differential refraction of neutrinos.

1st submission October 6, 2011 - latest revision October 12, 2011

## Introduction

Recently the OPERA [1] collaboration and the MINOS [2] collaboration reported evidence for the superluminal propagation of neutrino pulses over long baselines. The propagation speeds and are estimated at

$$v_\nu = c \times \left(1 + \begin{cases} 5.1 & \pm 2.9\,(\mathrm{stat+sys}) & \times 10^{-5} & \mathrm{MINOS} \\ 2.48 & \pm 0.28\,(\mathrm{stat}) \pm 0.30\,(\mathrm{sys}) & \times 10^{-5} & \mathrm{OPERA} \end{cases}\right)$$

where $c$ is the nominal speed of light. The superluminality is not significant for MINOS but is quite significant for OPERA. Both the NuMI beam used by MINOS and the CNGS beam used by OPERA are mostly muon neutrinos and travel over a distance of $\sim 730\,\mathrm{km}$ to the "far" neutrino detectors. The NuMI and CNGS beam have neutrino energy centered at $\sim 3\,\mathrm{GeV}$ and $\sim 17\,\mathrm{GeV}$ respectively.

Claims of superluminal propagation must be met with a great deal of skepticism, because this result is at variance with some of the foundations of modern physics, and raises problems with fundamental concepts like causality. This extraordinary result, if true, would require an extraordinary explanation. Here we catagorize a various explanations and point out a category that could be very easily tested.

One class of explanations is that these ultra-relativistic neutrinos actually do travel faster than light in most or all environments. This is in tension with the observations of lower energy neutrinos emitted from a supernova (SN-1987a), and observed by a variety of neutrino detectors on Earth [3,4,5], which must have traveled at $(1 \pm 2 \times 10^{-9}) \times c$. However since SN-1987a neutrinos are at much lower energy, $E_\nu \le 10\,\mathrm{MeV}$, there is no clear contradiction between SN-1987a and OPERA+MINOS since the speed may be energy dependent.

Another difference between OPERA+MINOS and SN1987a is that in the former cases the neutrinos traveled most of the distance through rock while in the latter case the neutrinos traveled through the very much less dense interstellar and intergalactic space. This could be the cause of the different apparent speeds of propagation. For the rest of this letter we use the term "dense matter" to refer to materials with densities of normal solids and liquids (roughly $\rho \gtrsim 1\,\mathrm{g/cm^3}$) and treat rock as if it were a generic form of dense matter. The expectation is that the chemical composition does not matter much.

Let us divide possible descriptive explanations into various options

1. high energy neutrinos always travel faster than $c$; or
2. high energy neutrinos travel faster than $c$ in dense matter, because
   a. of some interaction with the matter, or because
   b. distances inside matter are smaller than their exterior dimensions would suggest.

Options 1) and 2a) have all of the causality problems associated with superluminal motion, however option 2b) is really an issue of non-Euclidean geometry, something we are very familiar with from the general theory of relativity (GR). With the non-Euclidean option there need be no problem with causality because the motion is not superluminal, it only appears to be. The reason it appears to be superluminal while it is not is because we are using the wrong model of non-Euclidean geometry, *i.e.* GR is incomplete and missing an important component. Furthermore this non-Euclidean geometry would effect all particles not just neutrinos and have no energy dependence.

## Neutrino Refraction by Matter

We take it for granted that all particles including neutrinos behave as waves and for the moment we treat neutrino propagation through dense matter as if it were through a uniform medium. Waves are characterized by a frequency, $\nu$, and a wavelength, $\lambda$ which one can combine to form a phase velocity: $v_p = \lambda \nu$. Note $v_p$ is only a pattern speed while wave packets, which might describe a particle, travel at the group velocity, $v_g = v_p - \frac{dv_p}{d\ln\lambda}$. For option 2a) we must allow for $v_p > c$ while for option 2b) the actual phase velocity of neutrinos will always be $c$ or less, however if one defines $v_p^{\text{eff}}$ to be the effective phase velocity as determined by the exterior geometry then $v_p^{\text{eff}} \neq c$. Henceforth we will always use the "eff" notation for effective so as to be less prejudicial as to the true physical meaning of the various quantities we use.

In every case simple matching of the moving wavefronts at the interface of materials with different phase velocities tell us that the direction of propagation of the wave will change at the interface, and this change in direction is given by Snell's law [6]

$$\frac{\text{Sin}[\theta_{\text{ext}}]}{\text{Sin}[\theta_{\text{int}}]} = \frac{v_{p,\text{ext}}^{\text{eff}}}{v_{p,\text{int}}^{\text{eff}}}$$

where "int" and "ext" refer to inside and outside of dense matter, respectively, and the angles $\theta$ refer to the angle of incidence relative to the normal of the boundary of the material. Conventionally in optics one refers to the index of refraction, $n^{\text{eff}} = c/v_p^{\text{eff}}$, rather than the phase velocity.

We use an index of refraction appropriate to OPERA/CNGS which may or may not apply to the other beams such as NuMI which has a lower energy. In particular we assume that $v_{p,\text{ext}}^{\text{eff}} = c$ (referring to air or vacuum) while $v_{p,\text{int}}^{\text{eff}} - c \approx 2.5 \times 10^{-5}$ (referring to rock) corresponding to an index of refraction slightly less than unity. Note that light propagation in air has $c - v_{p,\text{int}}^{\text{eff}} \approx 3 \times 10^{-4} c$ which is only 10 times larger in magnitude and produces a variety of easily noticed phenomena such as mirages. One should expect that neutrino refraction required to explain OPERA+MINOS would also be easily noticed if it varied significantly with density.

With these assumptions Snell's law now becomes

$$\text{Sin}[\theta_{\text{int}}] = \frac{v_{p,\text{int}}^{\text{eff}}}{c} \text{Sin}[\theta_{\text{ext}}].$$

Since the index of refraction is less than unity we have *total external reflection* for grazing incidence neutrinos $\theta_{\text{ext}} > \theta_{\text{crit}}$ where

$$\theta_{\text{crit}} = \text{Sin}^{-1}\left[\frac{c}{v_{\phi,\text{int}}^{\text{eff}}}\right] \approx 90° \left(1 - \frac{2}{\pi}\sqrt{2\frac{v_{p,\text{int}}^{\text{eff}} - c}{c}}\right) \approx 89.6°$$

where we have used the OPERA value of $v_{p,\text{int}}^{\text{eff}}$. The bending angle of the refracted wave $\delta_{\text{refract}} \equiv \theta_{\text{int}} - \theta_{\text{ext}}$ varies from $0°$ for normal incidence ($\theta_{\text{ext}} = 0°$) to $\delta_{\text{refract}}^{\text{crit}} \approx 0.4°$ which occurs when $\theta_{\text{ext}} = \theta_{\text{crit}}$ at which point all the neutrinos are reflected rather than refracted. Note that $\delta_{\text{refract}}^{\text{crit}}$ is much larger it's mean for random incidence angles. For all incidence angles some neutrinos will be reflected (Fresnel reflection). The bending angle of the reflected wave is $\delta_{\text{reflect}} = 2(90° - \theta_{\text{ext}})$ varies from $180°$ for normal incidence waves to $0°$ for waves parallel to the surface. At the critical angle $\delta_{\text{reflect}}^{\text{crit}} = 2\delta_{\text{refract}}^{\text{crit}} \approx 0.8°$. The fraction of energy in the reflected wave, the Fresnel reflection coefficient, is spin dependent and requires a more complete theory for the cause of refraction to calculate. We may use unpolarized electro-magnetic waves as an illustrative expample: the Fresnel reflection coefficient varies from $\left(\frac{v_{p,\text{int}}^{\text{eff}} - c}{v_{p,\text{int}}^{\text{eff}} + c}\right)^2 \approx 1.5 \times 10^{-10}$ for normal incidence to 1 for $\theta_{\text{ext}} \geq \theta_{\text{crit}}$, with a very rapid rise as one approaches the critical angle [7].

In normal refraction the momentum transferred to allow for the bending of trajectories of particles (waves) comes from the refractive material. In GR light rays are curved by gravitational lensing which can be described by a varying index of refraction of space-time. In GR lensing the momentum transfer to the lens is given the gravitational tug on the lens by the passing photons. If option 2b) were correct then one might say that the momentum transfer is due to an additional gravitational force. If option 2a) were correct then some other kind of new force must be at work to transfer momentum. These forces are applied to the neutrinos at the interface where the wavefronts change direction. While the momentum transferred to the refractive material will depend in detail on the Fresnel reflection coefficient we expect it to be maximize near the critical where a fraction $2c\sqrt{v_{\phi,\text{int}}^{\text{eff}\,2} - c^2}\,\big/\,v_{\phi,\text{int}}^{\text{eff}\,2} \approx 0.014$ of the total momentum is tranferred.

## Probing Superluminal Neutrinos

We have seen that if $v_{p,\text{int}}^{\text{eff}} \neq c$ in dense matter then this implies the ability to lens neutrinos (change their direction of motion). Refraction/reflection can be used to test this phenomena. A single element lens inserted in the neutrino beam would deflect neutrinos up to $2\times 0.4°$ at the critical angle. If the angular dispersion of the neutrino beam were much smaller then this then such a lens could significantly defocus the neutrino beam greatly decreasing the neutrinos counts at the far detector but without significantly effecting the counts at a near detector (if present). It is also possible that inhomogeneities in rock along the beam could have significantly defocused the beam.

A far easier test would be to use the fact the neutrino beams are pulsed. This means that any lens put in the beam would experience a periodic force as the neutrinos deflected when the beam is on and not when the beam is off. The pulse frequency is typically in the sub-Hertz range and one might be able to detect this acoustically or mechanically. Since there is some deflection and reflection no matter what the incidence angle there should be an acoustic signal in anything in the neutrino beam, and could in principle have been noticed in routine operations of these experiments, however the maximum effect is for grazing incidence beams near the critical angle. To quantitatively test this scenario one would not need a complete lens that covered the entire beam. Since for total external reflection force is purely a surface effect one need not have a large volume for the partial lens but the lens could be constructed only of surfaces. Perhaps a simple metal plate attached to a piezo-electric microphone or even a stethoscope would work. The plate would act as something like a "gong" against which the neutrinos would "tap". This is illustrated in Figure 1.

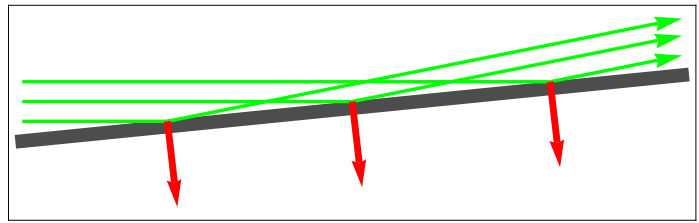

**Figure 1.** Grazing incidence neutrinos (green) undergo total external reflection off of refractive matter (gray) imparting momentum to the refactive matter (red). The angles are exaggerated here.

## NuMI Numerology

As an example we consider the NuMI neutrino beam which is used by MINOS [8]. Again note that we are using the OPERA/CNGS index of refraction which may or may not apply to the lower energy NuMI beam. The time structure of the NuMI beam produces 5 or 6 "batches" of neutrinos for every main injector "spill" which lasts $9.7\,\mu$s. The spills consist of $\sim 10^{13}$ protons which may produce a comparable number of neutrinos. The spills are separated by 2.2 s while the batches have duration $1.582\,\mu$s and are separated by $0.038\,\mu$s. Thus the temporal structure of the beam is roughly $9.8\,\mu$s on and 2.2 s off. Since the momentum transfer from lensing tracks the neutrino momentum flux this is also the time structure of the forcing function on any lens. The temporal spectrum would be roughly white noise from about 1 Hz up to about 100 kHz, so most of the acoustic energy is at the highest frequency. The time structure of the acoustic waves in the lens will be largely determined by the acoustic structure of the lens itself.

A lens surface at the critical angle, $\theta_{\text{crit}} \approx 89.6°$, will deflect all incident neutrinos by $2\times 0.4°$ and receive an impulse of 1.4 % of the total pulse momentum, which if there are $10^{13}$ neutrinos per spill, with mean energy 3 GeV, corresponds to a $\sim 20\,\text{g\,cm/s}$ impulse per spill, roughly equivalent to a tap of a human finger. This should be easily detectable. If the lens

$\theta_{\text{crit}} \approx 89.6°$         $0.4°$         $1.4\%$

$10^{13}$

surface does not intersect the entire beam this impulse should be scaled down by the fraction of the area intersected, and it will also scale with the intensity of the neutrino beam.

The CNGS neutrino beam [9] used by OPERA has a similar number neutrinos per beam spill, a somewhat complicated beam time structure and a cadence which is roughly three times longer: 6s versus 2s. Most imporantly the neutrino energy is roughly 3 times larger leading to a tripling of the available impulse from each spill for a single total external refraction. The T2K neutrino beam [10] produced at J-PARC has lower energy neutrions ($\sim 500$ MeV) but a larger flux ($\sim 10^{14}$ protons per spill) which yields a somewhat smaller signal than NuMI. The fact that T2K has a much shorter baseline to the far detector (295 km) is irrelevant for our purposes.

## Further Exploration

If this all worked one could go further and try to measure how $v_{p,\text{int}}^{\text{eff}}$ depends on neutrino energy and flavor as well as on the material the lens is made of. Density, isospin density, and chemical composition are obvious variables. Perhaps in the future one might also be able to determine whether electro-magnetic fields also have a lensing effect on neutrinos and how this depends on orientation of the fields? This is particularly interesting because electromagnetic fields have very different composition and stress-energy from normal matter.

If the effect is due to non-Euclidean geometry (option 2b) one would expect there to be no dependence on neutrino energy and flavor (up to very small terms having to do with the very small neutrino mass) so as to keep the equivalence principle.

## Other Practical Application

Lensing of neutrinos in this fashion, beyond it's implications for the foundations of physics, could have a number of practical applications. Obviously one could steer neutrino beams as well as focus them, perhaps even constructing neutrino storage rings. The acoustic/mechanical effects could lead to a different and perhaps easier way to detect neutrinos. Recall all of the neutrinos are lensed and contribute to the acoustic effects, whereas only a very small fraction of the incident neutrinos are detected in a traditional neutrino detector.

# What's Wrong With This Picture?

We have derived a number of fairly specific predictions for consequences of certain classes of models for superluminal neutrino propagation using a only a few assumptions. These are derived from very general principles having to do with wave dynamics. This begs the question of whether the models or the assumptions are self-consistent or consistent with other measurements. These sorts of scenarios are admittedly implausible by normal physics criteria, however superluminal neutrino propagation will require something unusual to explain it. Now we examine various problems and issues with this differential refraction scenario.

## Smoothing Needed

In the normal optical context the reason one speaks of refraction rather than scattering is because the scattering of light by individual atomic structures adds coherently in the forward scattering direction with nearly complete cancellation in other directions. This is true when $\lambda$ is much larger than the inter-atomic distance. However in the case of the OPERA neutrinos $\lambda_\nu \sim 10^{-14}$ cm which is thousands of times smaller than the inter-atomic distance scale and comparable to the size of an atomic nucleus. Another instance where refraction works is where the refracting medium is actually intrincially smooth. For the above analysis to work we require that the index of refraction be smooth enough that incoherent scattering off of inhomogeneities is small. We are supposing that this index depends on the density of something to do with matter, *e.g.* mass density, electron density, baryon density, isospin density, *etc*. In most solids (as well as liquids and gases) these quantities are uniform on a range of scales larger than the inter-molecular distance up to some granularity scale which depends on the medium. However the neutrino waves in question have wavelengths much smaller than the inter-molecular distance in rock and their propagation would be effected by variations in the index of refraction on molecular scales. On molecular scales the fractional density fluctuations in quantities like the electron density are of order unity, while the variations of baryon or isospin density are much greater than unity because nuclei are much smaller than atoms. Furthermore these fluctuations will vary randomly with time due to thermal motion on length scales much larger than $\lambda_\nu$. If we imagine that there are molecular scale fluctuations in the index of refraction with amplitude $\delta n^{\text{eff}} \sim 10^{-5}$ then scattering off of molecular scale inhomogeneities would add incoherently and large chunks of matter would be transluscent to neutrinos and would greatly dilute the neutrino flux in long baseline neutrino experiments (LBNEs). This is in contradiction with the phenomenology of LBNEs: which receive roughly the expected rate of neutrino events. Furthermore coherent scattering of individual multi-GeV neutrinos with molecular or atomic scale fluctuations would impart enough energy to disrupt the molecular and atomic bonds making neutrinos easily detectable. This is also in contradiction with established neutrino phenomenology. Such effects could have important consequences for supernova dynamics, depending on the energy dependence.

In order to circumvent this problem one would need to smooth the index of refraction over a large enough patch of matter so as to damp small scale structure in $\delta n^{\text{eff}}$ as well as distribute the momentum transfer from individual neutrinos over enough material so that it would not have been detected. This could be done with a very light force carrier or some other unspecified mechanism. No matter how this smoothing is accomplished we require that it be much larger than the atomic scales. If we denote the smoothing scale $\bar{L}$ we require that $\bar{L} \gg 10^{-8}$ cm where $\bar{L}$ specifies the width of the transition from vacuum behavior of the neutrinos so matter behavior. Note that Snell's law is unaffected by $\bar{L}$ so long as the length scales of the various pieces of matter are much larger than $\bar{L}$. We would expect a large $\bar{L}$ to increase the sharpness of the transition from near total refraction to total reflection. The results from the proposed acoustic tests for refraction should depend on $\bar{L}$ and thus $\bar{L}$ could be determined by these experiments. If $\bar{L}$ is comparable to or larger than the size of the experimental halls the proposed tests would be increasingly difficult as much of the refractive effect would come from the rock surrounding the halls thereby decreasing the momentum transfer to any apparatus. References [11,12] effectively propose a smoothing scale comparable or larger than the Earth's radius which would certainly reduce the signal to an unmeasureable level. However we expect significant refractive reflection should occur in the scenario proposed in reference [13].

For total external reflection $\bar{L}$ will determine the penetration depth of the evanescent wave [14] which is set up on the surface of the refractive material. It is the dynamics of the evanescent wave which transfers the momentum between the neutrinos and the matter and is an important manifestation of the (new) interactions causing the decreased index of refraction in matter. The evanescent wave should certainly be the subject of intense study if total external reflection is found. This penetration depth decreases from infinity to some finite value as one increases $\theta_{\text{ext}}$ from $\theta_{\text{crit}}$ to $90°$. On the practical side if the thickness of the refractive materal is not much larger than this penetration depth then reflection will not be total and some of the neutrinos would penetrate through the material, reducing the momentum transfered.

Note that option 1) where refraction is not environment dependent can be thought of as option 2) in the limit $\bar{L} \to \infty$. In this way we can think of options 1) and 2) in a unified framework. Clearly in this limit there would be no signal from the proposed experiments.

## Constraints on non-Euclidean Geometry

In option 2b) we suggested that the distances inside solids be somewhat smaller than their external dimensions. First note that the above argumentation that one needs smoothing to preserve refraction and prevent incoherent scattering applies in this case as well. So there must be a smoothing scale much larger than the atomic scale. Any non-Euclidean explanation would have some effect on the structure of solids, and in particular near the surface of a crystal this geometry would necessarily induce deformations (strains) on regular crystal structures. If $\bar{L}$ were large enough this might not be noticeable. There are a variety of conventional surface effects for crystals and this could be masked by these. If there were a hole in a solid into which one inserted a peg then if $\bar{L}$ were comparable or larger than the size of the hole, these strains would also effect the peg. To supply the energy associated with the strain in the peg one would expect there to be an additional repulsive force required to insert and an opposite attractive required to remove the peg. These could be noticeable in addition to the usual Van der Waals forces which have the opposite sign.

Since the non-Euclidean option, 2b), would effect all matter rather than just neutrinos, one could also imagine probing this effect by making precision measurements of the speed (phase velocity) of regular light as a function of the proximity of matter. This would be noticeable if $\bar{L}$ were large enough. Presumably there are very strong constraints on this already. Of course one regularly transmits light through solids and liquids however it seems unlikely that a shift of $\sim 10^{-5}$ in the index of refraction would be noticeable. One would expect some residual changes in the structure of individual atoms, transition levels, *etc*. which for simple atoms can be predicted quite presicely. Since atomic transitions are measured precisely in a large variety of terrestrial and astronomical enviroments one would imagine that any density dependence would have been noticed.

## Summary

Superluminal neutrino propagation, if true, requires some major reworking of our understanding of physics - and may lead us to abandon some of our assumptions of how the universe works. With this in mind we may want to step back to some of our most basic precepts. Here we work in the context of the idea that superluminal propagation occurs inside "dense matter" but not outside, which is suggested by the lack of superluminal propagation from SN1987a. If neutrinos propagate at the same speed both inside and outside of matter then the proposed experiments should detect nothing.

The basic assumptions we make use of are: that particles travel as waves, that momentum is conserved, and that momentum is related to the vector wavenumber of the wave. This is implicit in all of quantum field theory, but can apply to a larger set of matter models. In interpreting OPERA+MINOS we also make the substantive assumption that the phase velocity gives the speed of propagation of particles. If the group and phase velocities differ then the effects described here should still occur, but the size of the effect may be larger or smaller. Given our assumptions, simple matching of the wavefronts at the surface of dense matter requires Snell's law, the bending of direction of the wavenumber, and the splitting of the wave into reflected and refracted waves. Some of the momentum of the neutrinos is transferred to the dense matter, and this transferred momentum impulse can be detected through acoustic/mechanical means. The effect is maximized for grazing incidence beams about $0.4°$ from parallel to a surface for the OPERA superluminality. The momentum transferred is not small, $\gtrsim 10$ g cm/s/beam spill (from the entire beam), and should be easily detectable. Measurements to detect this could be done rapidly and inexpensively!

We have discussed ways in which such large effects should already have been observed but have reached no definitive conclusions as to what range of $\bar{L}$ is viable. The physical mechanism which determines $\bar{L}$ has been left open.

If this impulse is not found it would rule out $\bar{L} \lesssim 10$ m and argue for the idea that superluminality, if present, is not dependent on the material it passes through. It would certainly not entirely rule out superluminal neutrino propagation.

If the impulse is found this would not only lead to a much better understanding of the phenomena of superluminal neutrinos but also would point to new ways of manipulating neutrinos. This test makes use of the difference of neutrino speeds, or more specifically effective phase velocities, inside and outside of dense matter. It does not address the question of whether neutrinos are fast or photons are slow [15]; or whether there are any issues with causality. The physical mechanism for the different speeds of propagation would need much further exploration.

**ACKNOWLEDGMENTS:** The author is grateful for discussions with Jeter Hall, William Wester, and Aaron Chou. This work was supported by the DOE at Fermilab under Contract No. DE-AC02-07CH11359